# Mesoscale simulation of semiflexible chains. II. Evolution dynamics and stability of fiber bundle networks

Robert D. Groot

*Unilever Research Vlaardingen, PO box 114, 3130 AC Vlaardingen, The Netherlands*

Network formation of associative semiflexible fibers and mixtures of fibers and colloidal particles is simulated for the Johnson-Kendall-Roberts (JKR) model of elastic contacts, and a phase diagram in terms of particle elasticity and surface energy is presented. When fibers self-assemble they form a network for sufficiently large fiber-solvent surface energy. If the surface energy is *above* the value where single particles crystallize the adhesion forces drive diffusion-limited aggregation. Two mechanisms contribute to coarsening: non-associated chains joining existing bundles, and fiber bundles merging. Coarsening stops when the length of the network connections is roughly the persistence length, independent of surface energy.

If the surface energy is *below* the value where single particles crystallize, a network can still be formed but at a much slower (reaction limited) rate. Loose (liquid-like) assemblies between chains form when they happen to run more-or-less parallel. These assemblies grow by diffusion and aggregation and form a loose network, which sets in micro-phase separation, i.e. syneresis. Only when the clusters crystallize, the coarsening process stops. In this case the length of the network connections is larger than the persistence length of a single chain, and depends on the value of the surface energy. All networks of semiflexible homopolymers in this study show syneresis. Mixtures of fibers and colloid particles also form fiber bundle networks, but by choosing the colloid volume fraction sufficiently low, swelling gels are obtained. Applications of this model are in biological systems where fibers self-assemble into cell walls and bone tissue.

## I. INTRODUCTION

The phase behaviour and dynamics of fiber networks is an important problem in soft condensed matter, with many practical applications e.g. as viscosifiers in paint and foods. Fibers often self-organize into bundles that subsequently aggregate into networks.[1] The process of network forming by fiber bundles is also important in many biological systems. For instance, plant cell walls are networks of largely crystalline cellulose fibers held together by a mixture of polysaccharides, hemicelluloses and pectins.[2,3,4] Another important and even more complex example is bone tissue, which is initially formed by self-assembly of collagen microfibrils and subsequent co-crystallisation of hydroxyapatite,[1] but ultimately shows hierarchical organisation over seven levels.[5]

To improve our theoretical understanding of the formation of these and similar structures we need to reproduce them in simulation. A first step is to understand semiflexible chains *without* adhesive interaction. Various scaling relations are predicted for their dynamics and rheology, see e.g. Morse[6,7] for an overview. In general fibers can be characterised by their length $L$, diameter $d$ and their persistence length $L_p$, the distance over which the fiber orientation is correlated. A second key parameter is the fiber concentration $c$, the number of fibers per unit volume. Two important concentrations are usually defined: $c^*$ is the concentration where chains start to overlap and $c^{**}$ is the concentration above which the distance between neighbouring polymers becomes smaller than the persistence length. For $L >> L_p$ we have random coils. In this case for ideal polymers[6] $c^* = R_e^{-3} = (LL_p)^{-3/2}$ ($R_e$ is the polymer endpoint separation) and $c^{**} = L^{-1}L_p^{-2}$. For swollen polymers $R_e \sim L^\nu$, where $\nu \approx 3/5$ is the swelling exponent, hence $c^* \sim L^{-3\nu}$. The renormalization group result for the swelling exponent is[8] $\nu = 0.588 \pm 0.001$. Using a Flory argument, the chain size is obtained by minimizing the free energy per chain;[9] for a semiflexible chain this is $F/k_BT = R_e^2/LL_p + V(L/L_p)^2/R_e^3$, where $V$ is the excluded volume interaction between segments of length $L_p$ and diameter $d$.

*Assuming* an isotropic state we have $V \sim L_p^2 d$, and thus find $R_e^5 \sim L^3 L_p d$, hence $c^* \sim L^{-9/5} L_p^{-3/5} d^{-3/5}$.

In the other limit, $L << L_p$, we have stiff rods for which[6] $c^* = L^{-3}$ and $c^{**} = c^*(L_p/L)^{1/2} = L_p^{1/2} L^{-7/2}$. Onsager[10] showed that stiff rods order spontaneously above volume fraction $\phi_n = \frac{\pi}{4} c_n L d^2 = 4d/L$. This is the isotropic-nematic transition (see e.g. Dhont and Briels[11] for a recent discussion). Theory of rod suspension rheology[11,12] is therefore limited to $\phi \sim d/L$. Apart from these two limiting cases, theory has also been developed for entangled semiflexible fiber networks of general persistence length.[6,7,13,14]

Fibers with attractive interactions are more complex because associations and physical entanglements are both important for structure and dynamics. For this reason we often need to rely on simulations to gain insight into these systems. Notably, Bolhuis *et al*[15] studied hard spherocylinders with a square well attraction. They obtained a phase diagram with isotropic, nematic and smectic phases, including isotropic-isotropic phase coexistence if the range of attraction is sufficiently large. For small interaction range this coexistence disappears, and is replaced by isotropic-nematic coexistence. The phase diagram crucially depends upon the range of attraction and the ratio of rod length to its diameter.

Flexible associative polymers were first simulated by Groot and Agterof.[16] They obtained a phase diagram which shows under what conditions a gel is formed. The opposite limit of stiff rods with a *fixed* concentration of cross-linkers (small molecules that bind the fibers together) was studied by Zilman and Safran.[17] Their theory predicts when stiff, thin rods form a gel. Both groups come to the conclusion that a gel is formed when the polymer concentration and the association constant are sufficiently large, i.e. to the right of the (red) percolation line p in Figure 1. Furthermore, both for rod-like and for flexible polymers it was predicted that phase separation will occur for relatively low polymer concentration and for strong association, i.e. below the (black) binodal curve b in Figure 1. What is unique to rod-like fibers[17] is that above a certain concentration of rods and cross-linkers, the rods will align to form fiber bundles. These bundles subsequently form a network; this phase is located to the right of the (green) fiber bundle line f in Figure 1. Thus, the nematic transition observed for long spherocylinders acts as driver to coarsen the network.

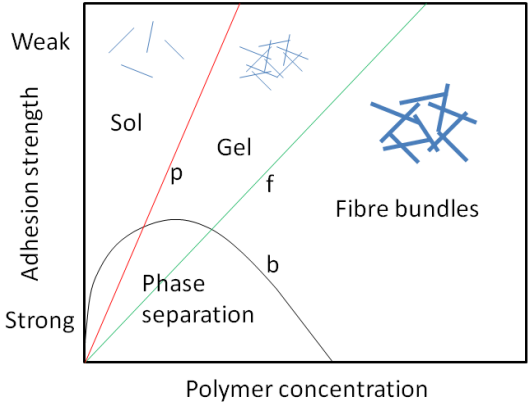

**Figure 1** Schematic phase diagram of associating polymers, after Groot and Agterof[16] (flexible polymers) and Zilman and Safran[17] (stiff rods). To the right of line p a gel is formed and below line b it phase separates. For stiff rods with cross-linkers, line f separates fiber gels from fiber bundle gels.

The area of interest here is neither the limit of flexible associative polymers, nor the limit of associative rods, but in between: semiflexible fibers that associate and form networks. For this system all the above comes together. Because the fibers associate to each other, the scaling relations for the rheology of semi-dilute semi-flexible polymers[6,7,13,14] cannot be applied, hence a number of key questions need to be answered by simulations. Some of these questions are: how does the strength of the network relate to fiber concentration, do semiflexible fibers also form fiber bundles, and if so what determines the thickness of the bundles? When these fibers form a network, what determines the mesh size, and under what conditions do these networks undergo syneresis? It is envisaged that a diagram as in Figure 1 applies, but the flexibility of the fibers introduces a new dimension into the problem.

For rod-like fibers of low concentration Dhont and Briels[11] argue that hydrodynamic interactions are not very important. However, for associative flexible fibers the early stages of network formation may be kinetically determined, in which case the pathway to form the network could be important. This again depends on hydrodynamic interactions. To investigate this point an efficient and reliable method has been developed to simulate the hydrodynamic interactions between fibers.[18] Using this method, the evolution of network formation is simulated. In the remainder of this work the simulation model is briefly reviewed in section II, simulation results for single particles and sticky fibers are described in section III, options to stop syneresis of fiber gels, including particle/fiber mixtures, are discussed in section IV, and a summary and conclusions are given in section V.

## II. SIMULATION MODEL

The simulation model used has been described in detail, and validated in a previous paper[19] (hereafter: Paper I), therefore only a brief outline is given here. Our starting point is the Fluid Particle Model (FPM) by Español.[20] Each particle $i$ is characterised by a position $\mathbf{r}_i$, a velocity $\mathbf{v}_i$, and a spin $\boldsymbol{\omega}_i$. The system is evolved by integrating the equations of motion: $\dot{\mathbf{r}}_i = \mathbf{v}_i$; $\dot{\mathbf{v}}_i = \mathbf{F}_i/m_i$; and $\dot{\boldsymbol{\omega}}_i = \boldsymbol{\tau}_i/I_i$, where $m_i$ and $I_i$ are the particle mass and moment of inertia.

Two types of interaction forces are used, conservative forces and dissipative forces. The dissipative forces consist of pair-wise radial and shear noise and friction.[18-20] The radial friction is characterised by a friction parameter $f$ and the shear friction is proportional to a parameter $\mu$. In the simulations described here we use $f = \mu = 1$. Details of the noise and friction functions used are given in Paper I.

For each pair of particles $i$ and $j$ we define the mean harmonic mean radius $R_{ij}$ and the mean diameter $d_{ij}$ as

$$R_{ij} = \frac{2R_iR_j}{R_i + R_j}; d_{ij} = R_i + R_j \qquad (1)$$

where $R_i$ is the radius of particle $i$. As conservative repulsive force we use the Hertz model. In this model the force between two elastic spheres is given by[21]

$$\mathbf{F}_{ij}^{Rep} = \begin{cases} E\sqrt{R_{ij}}(d_{ij} - r)^{3/2} \mathbf{e}_{ij} & \text{if } r < d_{ij} \\ 0 & \text{otherwise} \end{cases} \qquad (2)$$

where $\mathbf{e}_{ij} = (\mathbf{r}_i-\mathbf{r}_j)/|\mathbf{r}_i-\mathbf{r}_j|$. If $E$ is the elastic modulus of the material, the pre-factor should in fact be $^2/_3 E/(1-\nu^2)$ where $\nu$ is the Poisson ratio. In the present formulation this factor has been subsumed into the definition of the elasticity modulus $E$.

To model adhesive interaction between the particles we introduce surface energy. If the surface energy between particles and solvent is $\gamma$, the adhesion energy is $\Delta u = -2\pi\gamma b^2$, where $b$ is the radius of the contact zone. In the Hertz approximation we have $b^2 = R_{ij}(d_{ij}-r)$ hence the adhesion energy is $\Delta u = -2\pi\gamma R_{ij}(d_{ij}-r)$, and the adhesion force is

$$\mathbf{F}_{ij}^{Adh} = \begin{cases} -\varepsilon R_{ij} \mathbf{e}_{ij} & \text{if } r < d_{ij} \\ 0 & \text{otherwise} \end{cases} \qquad (3)$$

where $\varepsilon = 2\pi\gamma$. For generality $\varepsilon$ can take any value between any pair of particle types. This model is generally known as the Johnson-Kendall-Roberts (JKR) model, and has been validated experimentally to colloidal particle systems.[21]

To simulate semiflexible fibers, a harmonic spring force was introduced between neighbouring particles of a polymer chain:

$$\mathbf{F}_{ij}^{S} = -Cr_{ij} \mathbf{e}_{ij} \qquad (4)$$

and bending and torsion forces are introduced. Each particle $i$ has three internal unit orientation vectors ($\mathbf{a}_i$, $\mathbf{b}_i$, $\mathbf{c}_i$). The vector $\mathbf{a}_i$ acts as local director of the fiber. From the orientations of neighbouring particles $\mathbf{a}_i$ and $\mathbf{a}_j$ the torques and forces are calculated for a connecting beam of bending modulus $K_b$. The fibers thus obtained have persistence length

$$L_p = \frac{K_b}{kT} \qquad (5)$$

It has been checked by simulation that the correct endpoint distribution and (internal mode) dynamics is generated by this algorithm (see Paper I). To stabilize the spin temperature, a thermostating force is introduced that acts only on neighbouring particle spins. Details are given in Paper I.

In general the bending stiffness is given by $K_b = EI$, where $E$ is the Young's modulus of the material and $I$ is the area moment of inertia, $I = \pi d^4/64$ (this is different from the ordinary moment of inertia). For a isotropic, round fiber, we have

$$K_b = \pi E d^4 / 64 \qquad (6)$$

where $d$ is the fiber diameter. Torsional elasticity is characterised by torque $\tau = K_t \alpha/l$, where $\alpha$ is the rotation angle at the end of a clamped beam in radians, and $l$ is the length of the beam. The torsional stiffness is in general given by $K_t = GJ$, where $G$ is the shear modulus of the material and $J$ is the torsion constant of the sample. For round beams or fibers, $J$ is identical to the polar moment of inertia $J = \pi d^4/32$, thus for round fibers we have

$$K_t = \pi G d^4 / 32 = K_b /(1+\nu) \qquad (7)$$

At this point we used the relation $G = \frac{1}{2}E/(1+\nu)$, which holds for *isotropic* materials ($\nu$ is Poisson's ratio). Fiber bundles may not always be isotropic. In those cases, $K_t$ may be much smaller than $K_b$.

To include full hydrodynamics, solvent particles are added that behave as an ideal gas with mutual noise and friction forces. Long-range hydrodynamics between the colloid particles is generated if we impose the correct solid-fluid boundary conditions at the particle surface. These imply that the *relative* radial and transverse velocity of the solvent at the particle surface should vanish on average. This problem was solved recently by Groot;[18] this method was implemented here. Details are given in [18, 19].

## III. STICKY PARTICLES

To find sensible simulation parameters we first analyse the simulation model for simple spherical particles. We concentrate on particles of equal size $R_{ij} = R = d/2$. The interaction potential corresponding to the forces in Eq (2) and (3) is then given by

$$U(r) = \tfrac{1}{5}\sqrt{2}\, Ed^3 (1-r/d)^{5/2} - \tfrac{1}{2}\varepsilon d^2 (1-r/d) \qquad (8)$$

The association constant thus follows as[22]

$$K = 4\pi \int_0^d \left( \exp(-U(r)/kT) - \exp(-U_0(r)/kT) \right) r^2 \, dr$$

$$\approx 4\pi d^3 (x^*)^2 \sqrt{\frac{2\pi}{U''}}\, \left[ \exp(U^*/kT) - 1 \right]$$

$$(9)$$

where $U_0(r)$ is the first (repulsive) term in Eq (8), and $x^*$, $U^*$ and $U''$ are the relative position of the energy minimum, the energy minimum itself and its second derivative to $x = r/d$. These are given by

$$\begin{aligned} x^* &= 1 - \left(\varepsilon/Ed\sqrt{2}\right)^{2/3} \\ U^* &= \tfrac{3}{5}\varepsilon^{5/3} R^{4/3} E^{-2/3} \\ U'' &= 6\varepsilon^{1/3} R^{8/3} E^{2/3} \end{aligned} \qquad (10)$$

We can expect simple spheres to have a liquid or solid phase when the association constant is substantially higher than 1. For a narrow attractive interaction range ($x^* \to 1$) the adhesive hard sphere model can be used as a proxy. For this model the association constant at the critical point equals[23] $K_c = 9.24 \pm 0.04$, which for $E = 1000$ would imply $\varepsilon^{crit} \approx 58.3$. A more accurate estimate is obtained by using the adhesive *elastic* sphere model: spheres with a linearly increasing force and a short-range parabolic attractive force. The phase diagram for this model was given in Ref[23] in terms of the dimensionless temperature $kT/G$ and the interaction range $\delta$. The present model can be mapped onto this adhesive elastic sphere model by requiring that the mean-square difference of the Mayer functions of the two models is minimized. The Mayer function is defined as $M(r) = \exp(-U(r)/kT) - 1$; its space integral $\int M(r)\, d^3\mathbf{r}$ is the second virial coefficient.

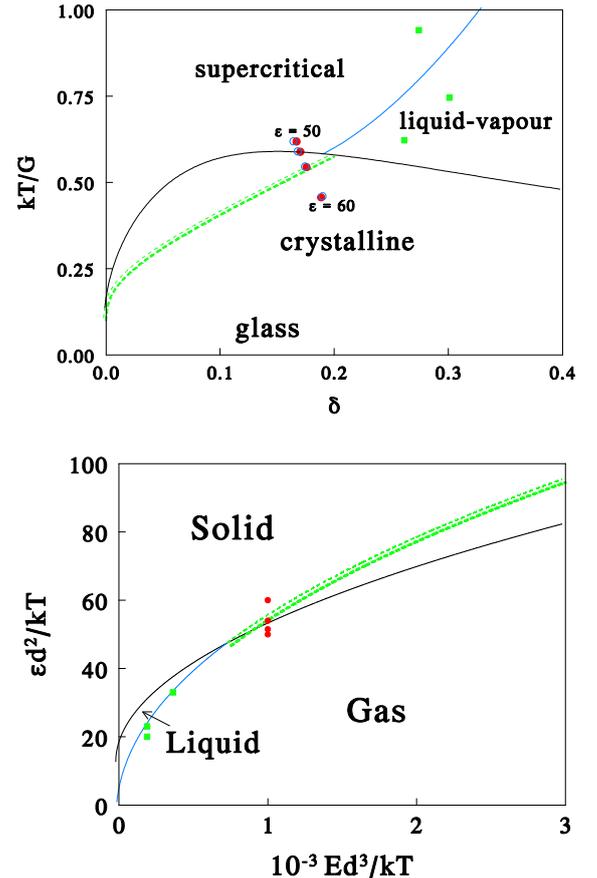

**Figure 2** Top: phase diagram of sticky elastic sphere model.[23] The red dots show the mapped position of the present JKR model for repulsion $E = 1000$ and $\varepsilon = 50$, 51.5, 54 and 60 based on Eq (12). Blue circles give numerical results from a Mayer function fit. The dashed curve is a line of hidden critical points. The green squares give the calculated positions of $E = 364$ and $\varepsilon = 33$, and for $E = 190$ and $\varepsilon = 23$ and 20 (supercritical for single particles). Bottom graph gives the same coexistence and critical curves, as function of elasticity and surface energy.

The interaction potential of the adhesive elastic sphere model (AES) is given by[23]

$$U^{AES}(r) = \begin{cases} \frac{3G}{\delta^2}(1-r/d_0)^2 - G & \text{if } r < d_0 \\ \frac{6G}{\delta^3}\left[\frac{1}{3}(1+\delta-r/d_0)^3 - \frac{1}{2}\delta(1+\delta-r/d_0)^2\right] & \text{if } d_0 < r \\ & < d_0(1+\delta) \end{cases} \quad (11)$$

where $G$ is the minimum energy, $d_0$ is the position of the energy minimum and $\delta$ is the range of the attractive potential. To map the JKR model to the AES model we can either fit the respective Mayer functions to each other as described above, or we can use an analytic approximation. In the analytic approximation we equate $G = U^*$; $d_0 = x^*d$, and we equate the second derivative of $U^{AES}$ in $r = d_0$ to $U''$ as given in Eq (10). This leads to an interaction range $\delta = (6G/U'')^{1/2}/x^*$, which indeed fits the Mayer function very well near its maximum. However, the interaction range $\delta$ obtained from the numerical fit (which takes the shape of the whole curve into account) is systematically smaller for $50 < \varepsilon < 70$. An excellent semi-empirical approximation for the parameter range studied here is obtained by removing the (dimensionless) factor $1/x^*$, i.e.

$$G = U^* = \tfrac{3}{5}\varepsilon^{5/3}R^{4/3}E^{-2/3}$$
$$\delta \approx \sqrt{6U^*/U''} = \sqrt{\tfrac{3}{5}}(\varepsilon/ER)^{2/3} \quad (12)$$

Note that $U^*$ and $U''$ both have dimension energy, and $\delta$ and $x$ are dimensionless. Conversely, from Eq (12) we can solve $\varepsilon$ and $E$ for given interaction range and energy minimum as

$$\varepsilon = \left(\tfrac{5}{3}\right)^{1/2}G\delta^{-1}R^{-2}$$
$$E = \left(\tfrac{3}{5}\right)^{1/4}G\delta^{-5/2}R^{-3} \quad (13)$$

Using the known phase diagram of the sticky elastic sphere model[23] this equation allows us to map out the phase diagram for the JKR model in Figure 2. The numerical and analytical approximations are shown in the phase diagram of the sticky elastic sphere model (left). In most examples shown $E = 1000$ and $\varepsilon$ is varied from $\varepsilon = 50$ to $\varepsilon = 60$. Special points are $\varepsilon^{melt} \approx 51.5$, the predicted melting transition point, and $\varepsilon^c \approx 54$, where an underlying critical point is predicted. The locations of some extra points in the liquid phase are also indicated. These parameters are used in Sec. IV C.

To test these predictions, simulations were done with 2000 particles of diameter $d = 1$, enclosed in a box of size $V = 9 \times 9 \times 40$, where all particles were lumped together in the left part of the system at the start. The time step used was $\delta t = 0.01\, d(m/k_BT)^{1/2}$, where $m$ is the particle mass. For $\varepsilon = 50$ the liquid phase immediately sublimated, for $\varepsilon_{ij} = 55$ and 60 a liquid-vapour coexistence appeared over a few thousand time steps, and for $\varepsilon_{ij} = 65$ the liquid phase crystallized. Bringing the interaction down to $\varepsilon = 60$ or 55 the solid did not melt, hence the solid phase is thermodynamically stable phase at $\varepsilon = 55$ and 60. Bringing the interaction further down to $\varepsilon = 51$, starting from a crystalline phase, the solid sublimated whereas it did not at $\varepsilon = 52$. Hence, the simulated melting point is $\varepsilon^m = 51.5 \pm 0.5$, which is spot on the predicted value based on the adhesive elastic sphere model. For all these systems the liquid phase is meta-stable, because the interaction range is too small to allow a stable liquid phase,[23] see Figure 2.

## IV. SEMIFLEXIBLE STICKY FIBERS

Two things are of paramount importance for the structuring properties of a fiber network. The first is the mesh size of the network $\xi$, the second is the life time $\tau$ of the connections. In general the strength of a polymer network is determined by the number of connections per unit of volume, hence the elasticity or storage modulus is proportional to

$$G' \propto \frac{1}{\xi^3} \quad (14)$$

The proportionality contains temperature and, for frozen connections, the elasticity of the bonds. As a rule of thumb the viscosity of the network is given by

$$\eta \approx G'\tau \propto \frac{\tau}{\xi^3} \quad (15)$$

For a dilute polymer solution $\tau$ is the rotational diffusion time of the polymers, but this relation also holds for finite shear rate and volume fraction, where the rotational diffusion time itself depends on polymer concentration and shear rate.[14] For this reason, both polymer dynamics and network structure are of key importance for viscosity and rheology. Therefore we now study in more detail what determines the mesh size and the polymer relaxation time.

First we estimate the mesh size for non-associating fibers. This estimate is relevant for the early stages of network forming, when the polymers have to diffuse over a mesh size to find the nearest neighbour. In this case the mesh size is determined by the distance between randomly oriented polymers. Within this length scale the monomer concentration is $\rho = N_\xi/\xi^3$, where $N_\xi$ is the number of monomer units within a length scale $\xi$. For fractal polymers we have the scaling relation $\xi = dN_\xi^\nu$, where $d$ is the monomer size and $\nu$ is the swelling exponent, hence $\rho = (\xi/d)^{1/\nu}/\xi^3$. Solving for the mesh size we find

$$\xi = d^{1/(1-3\nu)}\rho^{\nu/(1-3\nu)} = (d\rho)^{-1/2} \quad (16)$$

where the last step follows for rod-like polymers, for which $\nu = 1$. In the simulations described below we use polymers at 5% volume fraction. For that case this estimate gives for a correlation length $\xi = 3.25d$. In the *late* stages of evolution, when a network has been formed, the mesh size may well be proportional to the persistence length.

Henceforth, throughout the results sections, reduced units will be used, where the colloid particle diameter $d$ is chosen as the unit of length and $k_BT = 1$ as unit of energy. Thus, the unit of time is $t^* = d(m/k_BT)^{1/2}$, where $m$ is the particle mass.

### A. The influence of hydrodynamics and solvent

The problem of self-assembly of polymers is related to the problem of the coil to globule transition when an isolated swollen polymer is brought abruptly into bad solvent conditions, e.g. by a temperature change. This problem was first brought up by De Gennes,[24] and subsequently analysed numerically by Kuznetsov et al.[25] and analytically by Pitard.[26] De Gennes described the collapse of a polymer by successive stages of folding and buckling, where the polymer forms a sausage that grows thicker in time. The outcome of the latter work is that for a chain of length $N$ the typical time scale of collapse is proportional to $\tau_c \sim N$ when hydrodynamics is included. Without hydrodynamics the time scale in the first stages of collapse is proportional to $\tau_c \sim N^{4/3}$, and in the late stages it is proportional to $\tau_c \sim N^{5/3}$. In the problem of self-assembly of block copolymers Groot et al.[27] simulated the evolution of one system with full hydrodynamics using dissipative particle dynamics, and also simulated the same system without long-range hydrodynamics. Whereas the system with hydrodynamics evolved towards a hexagonal phase, the system without hydrodynamics did not, but remained in a metastable network state. These results show that long-range hydrodynamics in some cases speeds up the process of self-assembly, and in other cases can be essential for correct evolution.

To investigate the phase behaviour and dynamics of semiflexible fibers, $N_c = 64$ chains of $L = 40$ particles were simulated, with spring constant $C = 10$, bending stiffness $K_b = 10$ and torsional stiffness $K_t = 7.5$. The chains were simulated in a periodic box of size $V = 30 \times 30 \times 30$, hence the monomer concentration is $\rho = N_c L/V = 0.0948$. The volume fraction of this system is $\phi = 0.05$, much larger than the estimated overlap concentration (Paper I)[19] $\phi^* \approx 0.002$. This overlap concentration was estimated on the basis of the simulated endpoint distribution of an isolated chain.[19] In some cases, in Sec. IV C, the box size was doubled in each direction to $V = 60 \times 60 \times 60$ and the number of polymers was increased to $N_c = 512$ chains of $L = 40$ particles to study finite size effects.

To facilitate these larger simulations the importance of explicit solvent is studied here. If we leave out the solvent a considerable gain in simulation speed is obtained. However, as mentioned above, the solvent generates long-range hydrodynamics. Without solvent we only have short-range noise and friction up to a distance of one particle radius between colloid particle surfaces, although we do maintain conservation of momentum. To enable a comparison between the two methods we study the diffusion of non-associating polymers. What is relevant for the (early stage) dynamics of network formation is the diffusion constant on the length scale of the mesh size $\xi$. If the association between fibers is weak, this will be an important time scale for rheology. It sets the shortest relevant time scale in the system. It also defines the time scale at which strongly associating polymers touch and start to form a network.

In simulations *with* solvent the step size $\delta t = 0.025$ is used. This time step is larger than in Sec. III; it leads to some 1-2% artificial temperature increase, which is accepted to gain simulation speed. For full long-range hydrodynamic interaction, 16679 ideal gas particles were added that interact with each other via radial and shear noise and friction ($f = \mu = 1$). When they collide with a colloidal particle surface, noise and friction are replaced by a collision step where a new transverse velocity and particle spin are drawn from the correct distribution, conserving momentum and angular momentum.[18,19] In simulations *without* solvent the step size $\delta t = 0.01$ is used for a better temperature control. This typically gives an error in the temperature of 0.5%.

The diffusion of non-interacting polymers was simulated in Paper I, this data is reproduced here in Figure 3. What is shown is the monomer mean square displacement, averaged over all colloidal particles. The horizontal dashed lines correspond to $\delta R = \xi = 3.25d$ and $\delta R^2/3 = \xi^2$. From the green fit curve we find the intersection with the lower dashed line at $t_\xi \approx 37$. This is the time that the beads in the system are on average displaced by a distance $\xi$. This corresponds to the time that neighbouring polymers start to touch each other, i.e. the time where the network starts to form. At this point the initial diffusive behaviour ($\delta R^2 \sim t$) crosses over to the collective dynamics of wormlike chains[28] ($\delta R^2 \sim t^{0.72 \pm 0.02}$). At very short time scale ($t < 1$) the dynamics is slightly faster than diffusive ($\delta R^2 \sim t^{1.08}$), which is a remnant of polymer inertia.

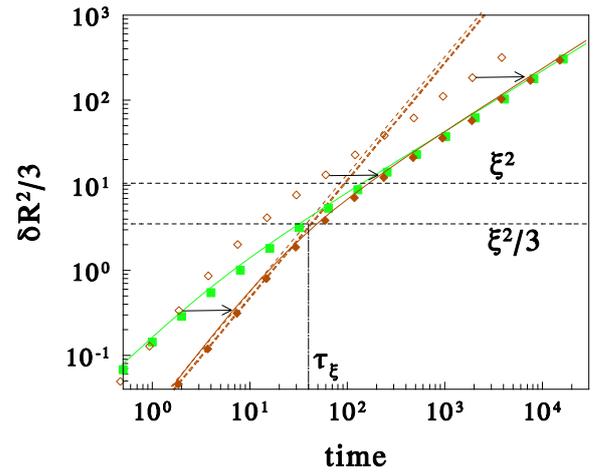

**Figure 3** Mean square monomer displacement for polymers of length $L = 40$ with solvent (green). The dashed black curves corresponds to displacement $\delta R^2/3 = \xi^2$ and $\delta R^2 = \xi^2$. The open brown symbols give the diffusion results for $L = 40$ chains without solvent; the full brown symbols give these results, time-shifted by a factor 3.7.

Also shown in Figure 3 is the (monomer) mean square displacement of polymers of the same length $L = 40$ and the same volume fraction $\phi = 0.05$, but *without* solvent included (brown symbols). Here we find a very clear inertial behaviour in the early stages for $t < 10$, $\delta R^2 \sim t^{1.37 \pm 0.02}$; i.e. a power law between ballistic ($\delta R^2 \sim t^2$) and diffusive ($\delta R^2 \sim t$) motion. Power law fits through the early and the late stage dynamics cross exactly at the distance scale $\delta R^2 \sim \xi^2$ that corresponds to the mesh size $\xi = 3.25d$. This is a measure of the mean free path of the polymers. For larger length scales (or time $t > 37$) the results with and without solvent superimpose if in the latter case time is rescaled by a factor 3.7.

To summarize, *with respect to free diffusion* a system with solvent can be mimicked by one *without* solvent. The solvent is important up to the time it takes for a polymer to diffuse over the distance to the nearest neighbour polymer. After that, the monomer mean-square-displacements with and without solvent superimpose if the simulated time scale in the solvent-free system is multiplied by a factor 3.7. The faster evolution without explicit solvent can be attributed to the lower number of friction centers.

## B. Network formation dynamics

Now that the dynamics of free polymers has been established, we can study the formation of polymer networks when the adhesive interaction is turned on. Apart from the adhesive interaction the same simulation parameters are used as in Sec. IV A. As expected, when $\varepsilon = 60$ micro-phase separation sets in, and a network is formed. In line with the prediction by Zilman and Safran[17] for rods, the semiflexible fiber system also forms a fiber-bundle network. A time series is shown in Figure 4. This shows a continuous coarsening of the structure, until about 400 time units. Thereafter only minor changes take place, where single fibers collapse on an existing bundle. It is observed that by $t = 50$ a few polymers have formed an initial fiber bundle network with a typical mesh size $\xi \approx L_p = 10$, which is then reinforced by other polymers. In all conformation graphs throughout this paper the *x*, *y* and *z*-axes are marked red, green and blue respectively.

If the same system of $\varepsilon = 60$ as in Figure 4 is simulated *without* solvent particles we find the evolution shown in Figure 5. In the first conformation, which has an evolution time $t = 25$, already quite some fiber bundles have formed. If evolution is primarily determined by diffusion, this state should compare roughly to time $t = 100$ of Figure 4, where solvent was included. The conformations are indeed similar. On longer time scales we see that the system is locked into one (frozen) conformation. Some minor changes are seen after $t = 200$ but it is not unreasonable to say that by this time the final state is reached. This suggests that statistically the same evolution is observed at roughly a four time faster rate because the friction coefficient of chains is lower without solvent.

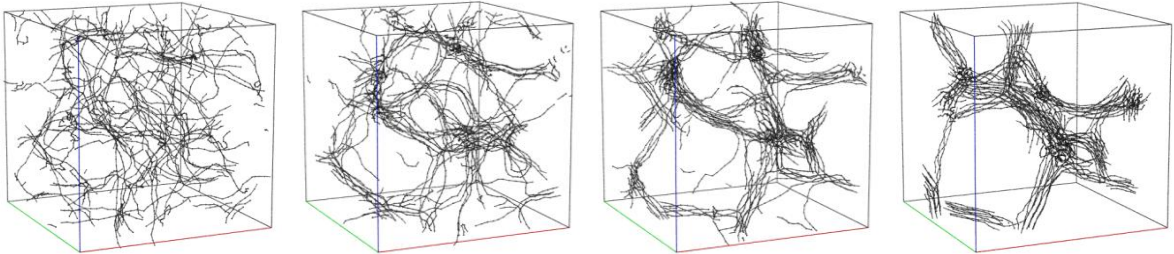

**Figure 4** Configuration of fiber-bundle network, formed by 64 chains of $L = 40$ particles with bending stiffness $K_b = 10$ and surface energy $\varepsilon = 60$ and solvent (full long-range hydrodynamics). The evolution time from left to right is $t = 25, 100, 400, 3200$. The *x*, *y* and *z*-axes are marked red, green and blue respectively.

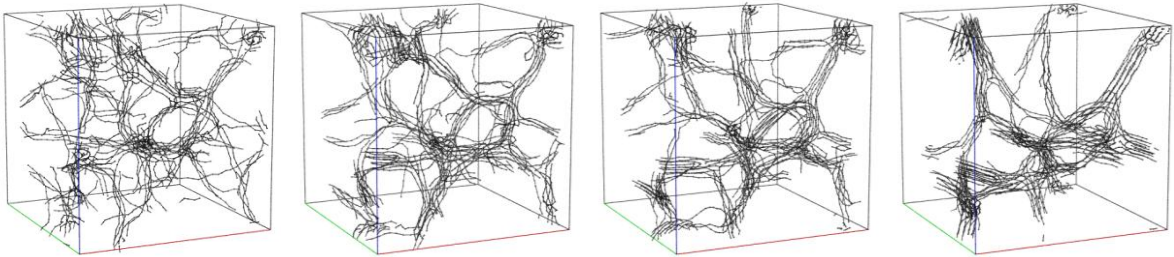

**Figure 5** Configuration of fiber-bundle network, formed by 64 chains of $L = 40$ particles with bending stiffness $K_b = 10$ and surface energy $\varepsilon = 60$ *without* solvent. The evolution time from left to right is $t = 25, 100, 400$ and $3200$.

When the interaction strength is reduced to $\varepsilon = 50$ and the evolution is followed from a fresh, disordered state, again a fiber-bundle network is formed. This is surprising, because the adhesive energy is *below* the crystallization transition for single particles. If the conformation of $\varepsilon = 50$ is used as starting point, and the adhesion parameter is reduced to $\varepsilon = 48$ or 47 the network of fiber bundles crystallises quickly, but at $\varepsilon = 46$ the network falls apart. This shows that the crystallisation transition for semi-flexible chains of $L_p = 10$ is $\varepsilon^m = 46.5 \pm 0.5$; i.e. a reduction by 10% as compared to single spheres.

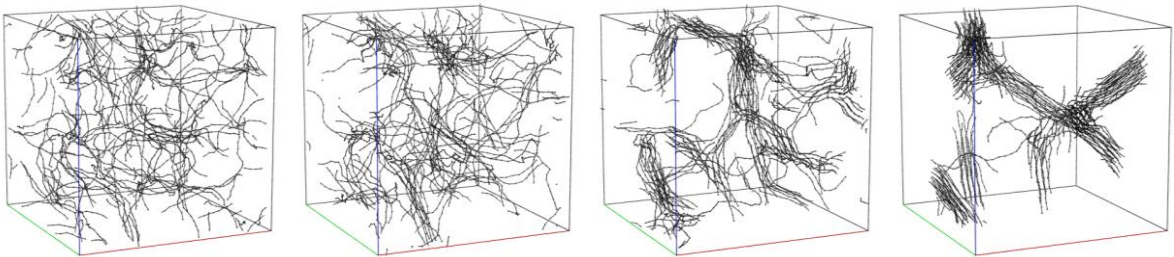

**Figure 6** Configuration of fiber-bundle network, formed by 64 chains of $L = 40$ particles with bending stiffness $K_b = 10$ and surface energy $\varepsilon = 50$ *with* solvent. The evolution time from left to right is $t = 100, 200, 800$, and $t = 3200$.

Figure 6 shows the evolution of a system of 64 chains of $L = 40$ particles with bending stiffness $K_b = 10$ and surface energy $\varepsilon = 50$. This shows that the system does form a network, but the rate of network formation is much slower than for $\varepsilon = 60$. Even at time $t = 100$ hardly any structure can be discerned, whereas for $\varepsilon = 60$ already a fiber bundle network had formed at this point (see Figure 4). However, the system at $\varepsilon = 50$ evolves to much coarser structures in the late stages of evolution. Whereas for $\varepsilon = 60$ bundles are formed of about length 10 ($\approx L_p$), for $\varepsilon = 50$ fiber bundles of roughly length 20 are seen. *Without* explicit solvent, evolution is again faster by roughly a factor four (see Figure 7).

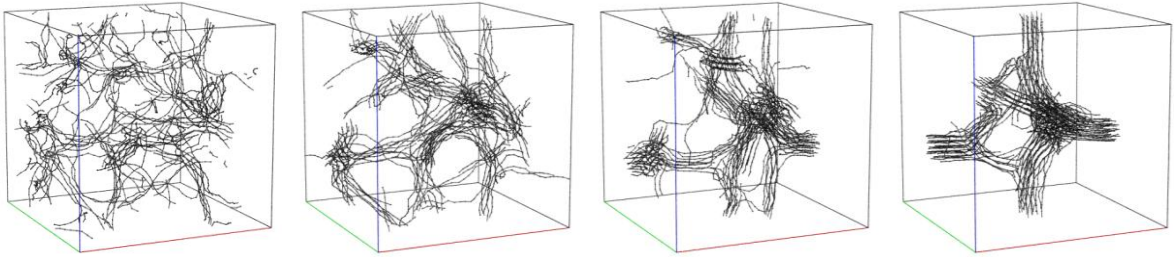

**Figure 7** Configuration of fiber-bundle network, formed by 64 chains of $L = 40$ particles with bending stiffness $K_b = 10$ and surface energy $\varepsilon = 50$ *without* solvent. The evolution time from left to right is $t = 50, 200, 800$, and $t = 3200$.

It is observed that at long time scale the $\varepsilon = 50$ system coarsens further than the $\varepsilon = 60$ system. This implies that the length of the network connections is *not limited* to the persistence length of the polymers. Therefore, *the hypothesis that the polymer persistence length always sets the mesh size of the network is incorrect*. This example shows that the mesh size *can* be larger than the persistence length of individual chains. When several chains self-assemble in a frozen fiber bundle, the collective persistence length of that bundle could however be larger than that of an individual chain if they bind very strongly, see Eq (6). In that case the persistence length is again linked to coarsening, yet also the coarsening is observed to stop. So if it is not the persistence length of the individual chains that sets the mesh size, the key question is, what *does* stop coarsening? Is it polymer crystallisation, or a balance of forces, or is this an artefact of the periodic boundary conditions? To investigate these points larger systems were simulated.

## C. What arrests coarsening?

In the $\varepsilon = 50$ system the mesh size appears to be limited by finite size effects (see Figure 6 and Figure 7). This is particularly visible when no explicit solvent is present, because this system is effectively four times as old as the system depicted in Figure 6. We may therefore wonder if the late stages of coarsening are arrested because of finite size effects, or because of an internal mechanism. To obtain a more accurate picture the system size was doubled in each direction and the number of polymers was increased by a factor 8 for both systems. These results are shown in Figure 8. In this system the fiber bundle length at $\varepsilon = 50$ is roughly 20 at $t = 3200$. These bundles are observed to crystallize around $t = 200$. For $\varepsilon = 60$ crystallization occurs earlier, around $t = 70$.

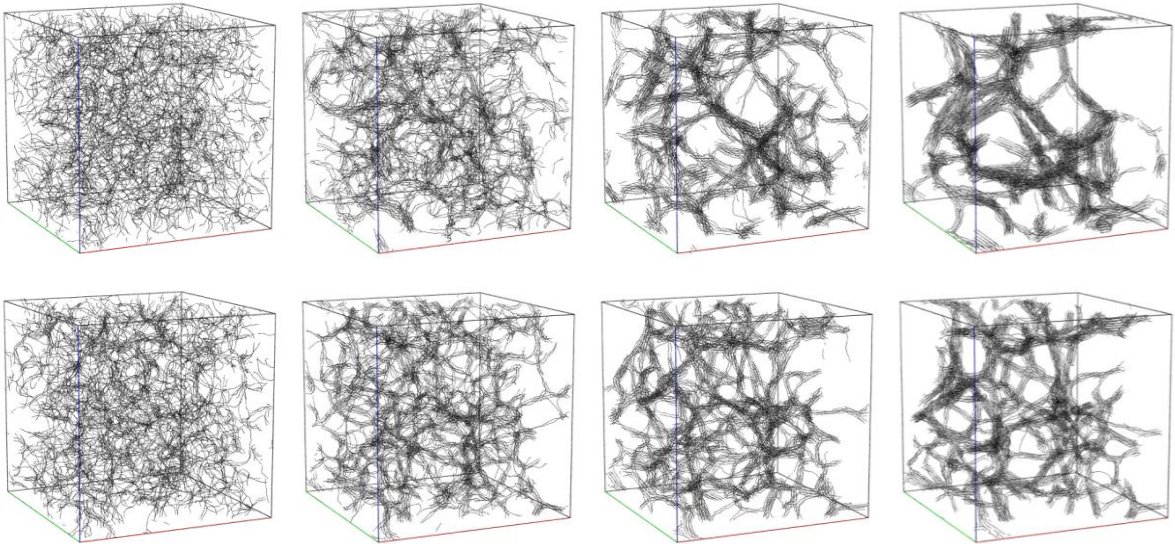

**Figure 8** Configuration of fiber-bundle network, formed by 512 chains of $L = 40$ particles *without* solvent, with bending stiffness $K_b = 10$ and surface energy $\varepsilon = 50$ (top row) and for $\varepsilon = 60$ (bottom row). The evolution time from left to right is $t = 10, 50, 400$ and $3200$.

To characterise the evolution of the systems the pressure was followed in time for three systems as shown in Figure 9. We find a slight pressure increase until $t \approx 10$; then it decreases steadily until $t \approx 200$, and finally it rises slightly until $t \approx 1000$. The interpretation is that in the initial stage the system forms a connected network. Once this is formed, it further coarsens by a mechanism where fiber bundles merge. At one point in time the process of coarsening stops and a minimum pressure is obtained. We may call this the coarsening time. By small rearrangements after the coarsening time, stress is released and loose fibers join existing bundles; and around $t = 1000$ the system enters a state where effectively nothing happens. The noise spectrum at this point has been analysed; we do not a find a power law distribution for the absolute difference between successive pressures measurements. This indicates that the system is *not* in a glassy state.

Two possible mechanisms might determine the coarsening time. Firstly, the system visually crystallizes near $t = 200$, which coincides with the coarsening time; hence crystallisation could be a mechanism to arrest coarsening. Alternatively, when fiber bundles merge the bundles are stretched, which increases tension. This tension will counteract the force that drives fiber bundle coalescence. At a given amount of coarsening, the energy gain obtained by an increased number of contacts will balance the penalty of a higher tension. To check which of these two effects is the main cause determining the final state, a third system with higher adhesion force $\varepsilon = 70$ was evolved over 3200 time units. The potential energy minima of the three systems are $U^*/kT = 1.62$ ($\varepsilon = 50$), $U^*/kT = 2.19$ ($\varepsilon = 60$) and $U^*/kT = 2.83$ ($\varepsilon = 70$). The pressure evolution of all three systems is shown in Figure 9.

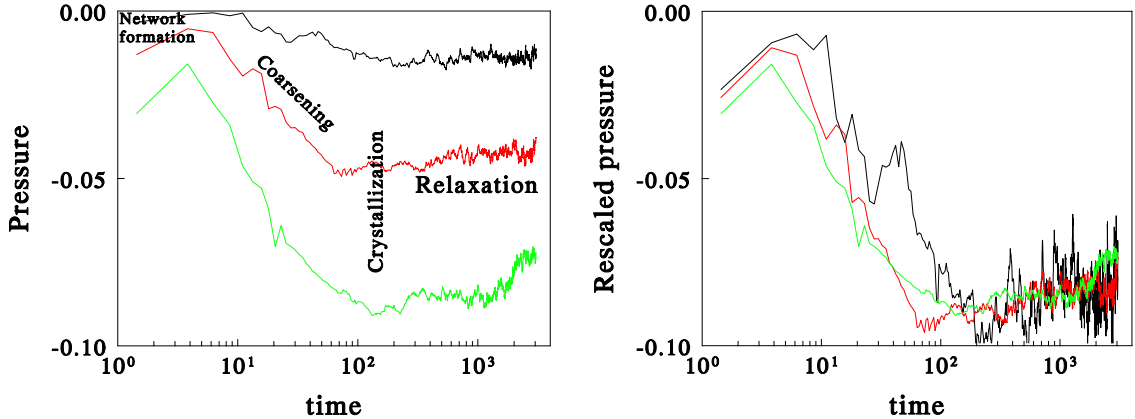

**Figure 9** Pressure evolution of fiber-bundle network for $\varepsilon = 50$ (black curve), $\varepsilon = 60$ (red curve) and $\varepsilon = 70$ (green curve). Right hand graph shows pressure, rescaled to match the $\varepsilon = 70$ results for $200 < t < 1200$.

In all cases the pressure stops declining at roughly the same point in time, although the $\varepsilon = 50$ system appears systematically slower than the other two. To aid comparison the pressures of the three systems were scaled to match the $\varepsilon = 70$ result. If the pressure minimum was caused by a balance of adhesive and tension forces, one might expect a different coarsening time for all systems. This is an argument in favour of crystallisation as mechanism to stop coarsening, but it is not conclusive.

To determine which effect is dominant the $\varepsilon = 70$ system at time 3200 was further evolved where the system size was adapted in small steps to steer towards a constant (vanishing) pressure. The resulting pressure and particle volume fraction are shown in Figure 10. The pressure cannot be kept constant but runs away to ever more negative values while the volume fraction keeps increasing. A similar behaviour is obtained for $\varepsilon = 50$. This indicates a mechanical instability that is not stopped by the (clearly visible) crystallinity of the connections. Instead, the crystalline bundles fold and buckle to minimize adhesion energy, much in the way as suggested by De Gennes.[24] This suggests that crystallinity is *not* the main factor to stop coarsening, or not the only one. Given the opportunity to contract, the system will do so, even if it's crystalline.

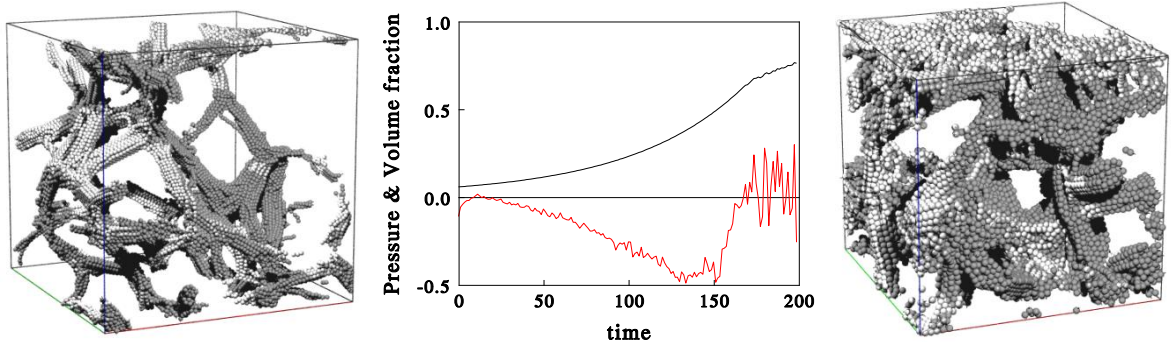

**Figure 10** Evolution of volume fraction (black curve) and pressure (red curve) for $\varepsilon = 70$ at varying system size. Left and right pictures show frozen cross-links at $t = 0$ and at $t = 100$. Time is counted from the moment the NPT simulation is started.

So what happens if the periodic boundary conditions are removed? The above results would predict a continued syneresis for isolated polymer gels. A first system was simulated with periodic boundary conditions in the *y*- and *z*-directions only. A gap in the *x*-direction allows the gel to adjust to an open end in one direction. In this system 20480 particles were inserted in a box of size 60×60×60, with walls at $x = 0$ and $x = 60$. A small wall association constant was used to glue the polymers to the wall and prevent desorption from the walls. Next, the gel was allowed to form over 200 time units, the time where crystallisation sets in, using adhesion $\varepsilon = 50$. At that point in time the walls were removed and the box was enlarged to $x = 80$. The conformation was then evolved over 2000 time units. Three snapshots are shown in Figure 11. This clearly shows the gel contracting in time for zero external pressure.

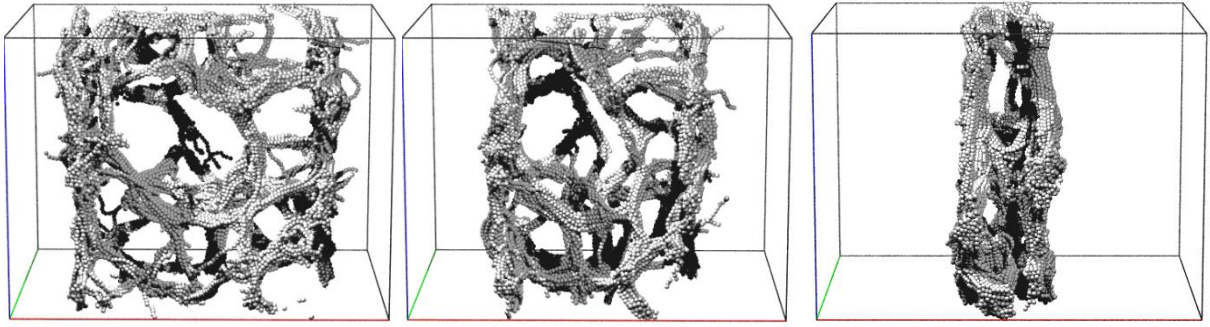

**Figure 11** Evolution of open system in one direction for ε = 50. The evolution times after the gel was released are t = 50, 200 and 2000. The *x*, *y* and *z*-axes are marked red, green and blue respectively.

A second system was completely immersed in vacuum, allowing the gel to contract in all three directions, see Figure 12. In this case, six walls were inserted and the polymers were allowed to equilibrate without adhesion forces to each other or to the walls. Next, the walls were taken away and the system was enlarged to 80×80×80. In this case fiber bundles at the corners of the gel are bent. This may generate a surface stress that tends to swell the gel. Nevertheless the gel folds and buckles and keeps contracting, although some holes remain visible at *t* = 2000. The continued syneresis shown in these simulations is also found in many practical systems. The folding and buckling observed here is very much in line with the mechanism proposed by De Gennes.[24]

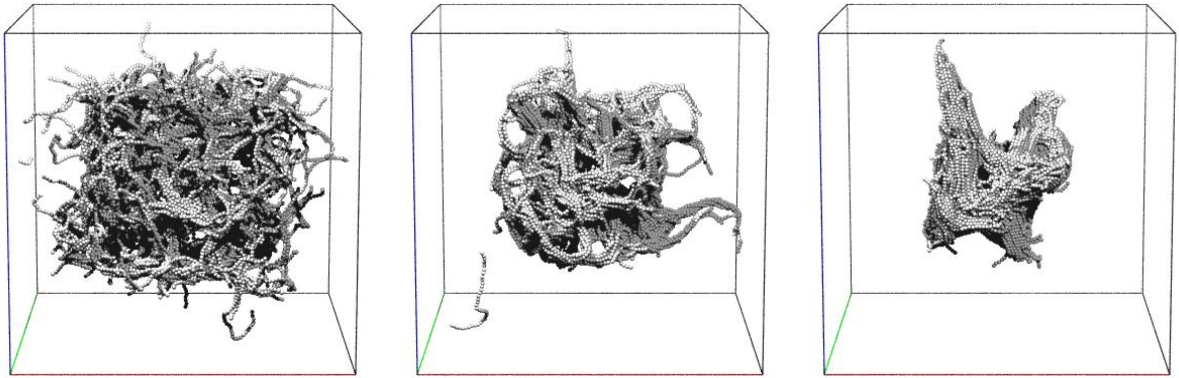

**Figure 12** Evolution of fully open system for ε = 50. The evolution times after the gel was released are t = 50, 200 and 2000.

As a final test, we check what happens when the connections *do not* crystallize. Two different parameter sets were chosen that correspond to the liquid state for monomers, and one point outside the liquid-vapour coexistence region (see Figure 2). For δ = 0.25 and *G* = 1.616 (the energy minimum for ε = 50 and *E* = 1000) we find ε ≈ 33 and *E* ≈ 364; and for δ = 0.3 and *G* = 1.333 we obtain ε ≈ 23 and *E* ≈ 190. These points are shown in Figure 2 by the green squares. Systems were prepared containing 512 polymers of 40 beads in a box of size 60×60×60, the same as for the systems shown in Figure 8. The resulting pressure evolution is shown in Figure 13, together with snapshots of the initial and final states for liquid connections.

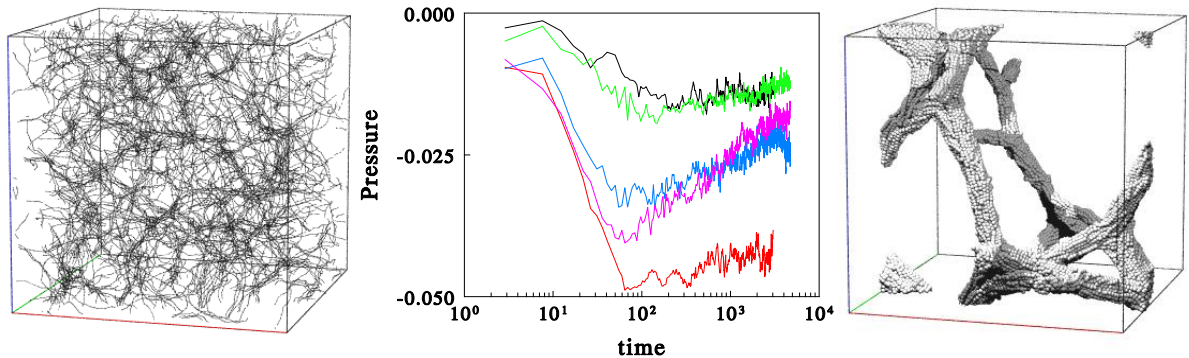

**Figure 13** Pressure evolution of solid fiber-bundle networks for ε = 50 (black) and ε = 60 (red), for "liquid" networks at ε = 33, *E* = 364 (blue); for ε = 23; *E* = 190 (pink) and for ε = 20; *E* = 190 (green). The left and right pictures give the configuration of the "pink" system at *t* = 10 and at *t* = 5000.

The network connections in these systems are liquid up to *t* = 4000. The pressure shows a gradual increase in time, particularly in the softer system (ε = 23; *E* = 190). In this system it was observed that the polymers reptate along the fiber bundle and join the larger domains. Thus, narrow bundles break, and on the time scale studied pressure increases as *P* ∝ $a+b\ln(t)$. This implies that $\partial P/\partial t \propto b/t \propto \exp(-P/b)$, which suggests a stress-induced break-up mechanism. In fact liquid tubes are known to break up spontaneously due to surface tension. This instability is known as the Plateau-Rayleigh instability.[29] In that case the time scale of break-up is determined by hydrodynamics. After $t = 4000$ the pressure appears to level off; at this point the system appears to crystallize to a soft solid and a slower diffusion-limited mechanism takes over. Because the polymer length is much longer than the size of the connections it is anticipated that the time to coarsen is proportional to the reptation time of the polymers. In a homogeneous bulk solution of concentration $\phi = 0.05$ for polymers of length 40, this is of order[19] $\tau_R = 25{,}000$. Here we have reptation in a soft solid, which is substantially slower.

To investigate the polymer mobility in the network connections, a simulation of 16 polymers of $L = 40$ beads was done for a free standing tube conformation. The system size was $10\times10\times24$. The tube converged to about 5 polymers in diameter, oriented in the $z$-direction. For parameters $\varepsilon = 23$ and $E = 190$ it showed diffusive behaviour over 5,242,880 steps of $\delta t = 0.01$, although the system did in fact order into crystalline domains. The obtained diffusion constant along the tube direction is, averaged over three subsequent runs, $D = (3.7\pm0.1)\times10^{-4}$. This diffusion constant implies a reptation time $\tau_R = 1.1\times10^6$, hence the actual time scale for the gel to break up by polymer diffusion is much larger than the time scale of the simulation shown in Figure 13. To prevent crystallisation completely, the adhesion parameter was further reduced to $\varepsilon = 20$. The diffusion constant along the tube in this case is $D = (8.7\pm0.1)\times10^{-4}$; the tube remains liquid and shows the characteristic undulations of the liquid phase. In a simulation of 512 polymers of 40 beads in a box of size $60\times60\times60$ a liquid network is formed that shows the characteristic pressure minimum (see green curve in Figure 13). In this case the pressure keeps increasing as $P \propto a+b\ln(t)$.

To summarize, the conclusions that we can draw from these simulations are 1) crystallisation is *not* the cause that stops the initial pressure drop, but 2) *without* crystallisation the network will keep on coarsening. If crystallisation is not the main cause to stop the pressure drop, the minimum pressure must be determined by a balance between the energy gain of fiber-fiber adhesion and the stress caused by stretching polymers to form a connection. Although crystallisation does not stop the initial pressure drop, it does stop the coarsening process of the network if periodic boundary conditions are imposed. Two coarsening mechanisms are observed when all polymers are associated, fiber bundles merging and individual polymers diffusing along the bundles, away from narrow connections.

## D. What determines the early stages of evolution?

The question remains why the $\varepsilon = 50$ system (shown in Figure 8) is so much coarser than for systems with stronger adhesion. To answer this question we first need to quantify the mesh size of the networks. Therefore the pair correlation is determined. The number of beads associated to any other in excess of a homogeneous distribution is given by $N(r) = 2\pi\rho r^2[g_2(r)-1]$, where $\rho$ is the monomer concentration. This function is shown in Figure 14 for the three systems $\varepsilon = 50$, 60 and 70. The pair correlations are fitted to a parabola, $N(r) \approx N_0(1-r^2/\xi_b^2)$. From this fit the thickness $d_b$ of the bundles is obtained as $d_b \sim (4N_0/\pi)^{1/2}$ and $\xi_b$ is *proportional* to the length of the fiber bundles. This shows that the $\varepsilon = 60$ and 70 systems are equal to within the noise ($\xi_b \approx 14$, $d_b \approx 4.5$), but the $\varepsilon = 50$ is significantly different ($\xi_b \approx 21$, $d_b \approx 6.0$).

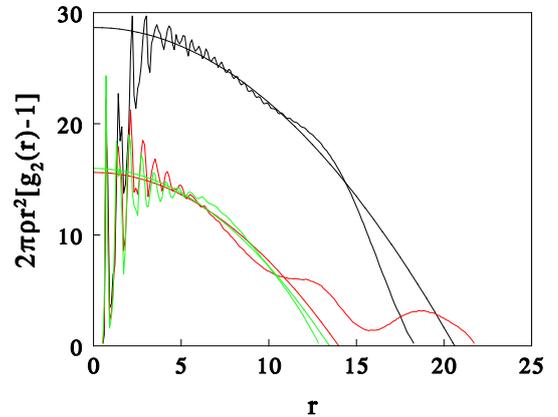

**Figure 14** Pair correlation of fiber-bundle network for $\varepsilon = 50$ (black curve), $\varepsilon = 60$ (red curve) and $\varepsilon = 70$ (green curve).

The early stages of network formation appear to depend on the association energy. For strong interaction ($\varepsilon = 60$ or 70) the process of network formation is initiated by the formation of dimers, where two chains associate to each other and form a complex. This initial stage is shown at the top of Figure 15, which gives the conformation of the $\varepsilon = 60$ system at $t_\xi = 37$ with full hydrodynamic interaction. This is the time where polymers have moved over a mesh size $\xi$. Some first tenuous connections are formed between neighbouring fibers. In this very early stage, neighbouring chains collapse onto each other and reorient to form parallel dimers. However, in this stage there are still many non-associated chains.

Thus, for $\varepsilon = 60$ (which corresponds to an energy minimum $U^*/kT = 2.2$, see Eq (10)) we have diffusion-limited aggregation where fiber bundles are formed immediately when two chains collide, and these fiber bundles subsequently aggregate and coarsen into a fiber bundle network. For $\varepsilon = 60$ we calculate the association constant between individual beads as $K_a = 13.4$. The adhesion energy is above the melting point for monomers $\varepsilon^m = 51.5\pm0.5$. On the other hand, for $\varepsilon = 50$ ($U^*/kT = 1.6$; $K_a = 7.1$) far fewer associations are formed, see bottom of Figure 15. Here we have reaction-limited aggregation, most probably because the adhesion energy is below the melting point for monomers. This allows local reorganisation of chains after they have formed a (loose) aggregate.

For $\varepsilon = 60$ and higher adhesion energy, reorganisation cannot take place because the strong adhesion effectively blocks dissociation once two chains have found each other. Once they meet they are stuck. The two chains align and a chain of associations zips the fibers together. In this case coarsening occurs mainly by the merging of several fiber bundles. In this case the length of the fiber bundles is set by the persistence length, the distance over which the chains are straight. In general, the bending energy of a chain of length $L$ and radius of curvature $R_c$ is $U_b = \frac{1}{2}K_b L/R_c^2 = \frac{1}{2}kT\,LL_p/R_c^2$. If all curvature is concentrated in domains of size $L \sim R_c \sim 1$, the bending energy in these domains is $U_b \sim \frac{1}{2}kTL_p$ and the bending energy of the connecting bundles vanishes. The mean bending energy per bead cannot exceed $kT$, hence the length of the connections $\xi_b$ must roughly equal the persistence length.

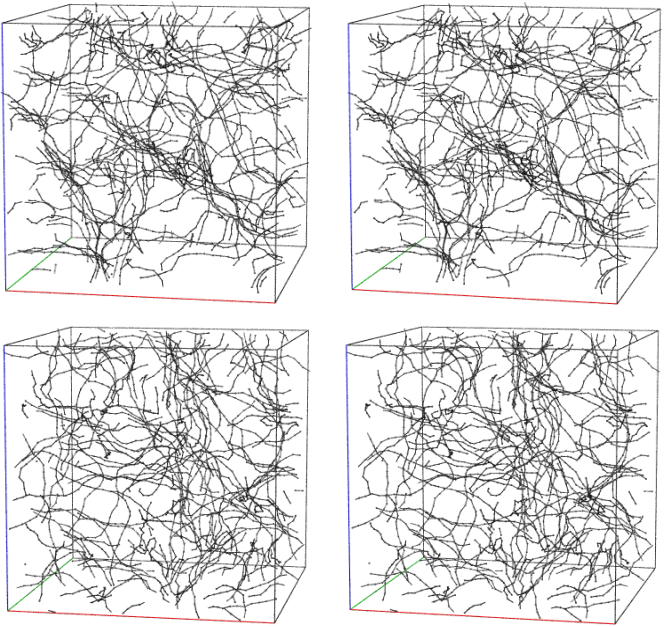

**Figure 15** Stereo pictures of sticky semiflexible fibers with full hydrodynamics at $t = t_\xi = 37$ – the hydrodynamic screening time – for $\varepsilon = 60$ (top) and for $\varepsilon = 50$ (bottom).

To summarize, semiflexible polymers melt – or rather sublimate – at a higher temperature (or lower surface energy $\varepsilon^p$) than monomers, $\varepsilon^p < \varepsilon^m$. For surface energy $\varepsilon^p < \varepsilon < \varepsilon^m$ the individual monomers do not stick to each other immediately on collision. Loose (liquid-like) assemblies between chains are formed when they happen to run more-or-less parallel. These assemblies grow by diffusion and aggregation and form a loose network, which sets in micro-phase separation. Only when the clusters finally crystallize, the coarsening process is stopped. For adhesion parameter $\varepsilon > \varepsilon^m$, however, the individual beads stick on collision. The adhesion forces are large enough to alter the local direction of the fibers, therefore aligned dimers and trimers are formed very early on. Once two chains find each other the dimer zips tight. This process occurs in many places simultaneously, and one chain can be member of several fiber bundles, so that a network is formed immediately. This network coarsens because non-associated chains join existing bundles, and because fiber bundles merge. Coarsening reduces the chain curvature, but this process stops when the curvature is less than the inverse persistence length because the bending energy per bead becomes or order $kT$. Therefore, in this case the length of the connections is roughly the persistence length.

## V. PARTICLE-FIBER MIXTURES

An option to stop segregation could be to use a mixture of two components: fiber A and colloidal particles B, where each of the components disperses well in the solvent, but where A and B associate to each other. This is the system considered in the theory by Zilman and Safran.[17] Within the present scope only a limited number of examples is investigated. We choose A as a semiflexible polymer of 40 beads and persistence length $L_p = 10$, as in Sec. IV A, and take B as a single (colloidal) particle. All repulsions are $E_{AA} = E_{AB} = E_{BB} = 1000$, and the adhesive interactions are chosen as $\varepsilon_{AA} = \varepsilon_{BB} = 0$ and $\varepsilon_{AB} = 80$. For this study we use $N_A = 2560$ fiber particles forming 64 A chains in a box of size $V = 30 \times 30 \times 30$ and use step size $\delta t = 0.01$, and vary the number of B particles as $N_B = 160, 320, 640, 1280$ and $2560$. In the last case the A and B particles are present in the ratio 1:1. These B particles serve as cross-linkers that will connect the A fibers together.

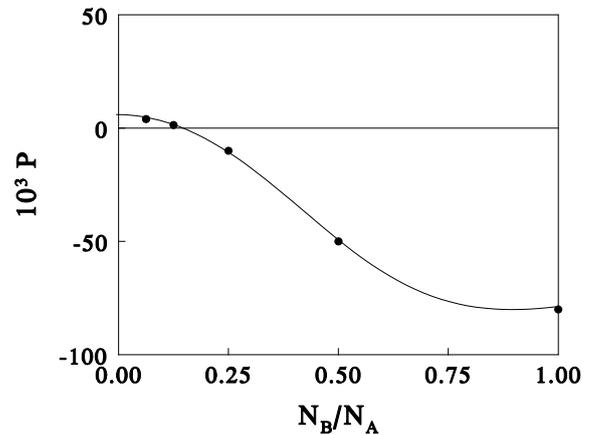

**Figure 16** Osmotic pressure in mixed fiber/particle system. The pressure vanishes at $N_B/N_A \approx 1/7$. The curve is merely a guide to the eye.

The addition of cross-linkers to a sample of A fibers directly leads to a pressure reduction, indicating phase instability. The B particles and A fibers in fact form crystalline structures, similar to the fiber bundles of stiff rods as predicted by Zilman and Safran.[17] For $\varepsilon_{AB} = 60$ no obvious associations are formed, for $\varepsilon_{AB} = 70$ some fiber bundles are formed, but these are in equilibrium with a relatively high concentration of non-associated B particles. For $\varepsilon_{AB} = 80$ nearly all B-particles are associated. In simulations where the fibers were confined by a semi-permeable membrane, the equilibrium concentration of B particles in the fiber-free area was 0.0083 for $N_B = 2560$ and 0.00022 for $N_B = 1280$. This means that the osmotic pressure of the mixed system is very close to the simulated pressure. For this reason $\varepsilon_{AB} = 80$ was selected for further study. The (osmotic) pressure is shown in Figure 16. This shows that the system will show syneresis if the cross-linker to fiber ratio $N_B:N_A$ is larger than $N_B:N_A \approx 1:7$.

Indeed, the systems $N_B:N_A = 1:16$ and $N_B:N_A = 1:8$ have a positive osmotic pressure. The question is thus, are these systems in a gel state or in a sol state? To answer this question a shear deformation $\gamma = 1$ was applied to these two systems using the Lees-Edwards boundary condition.[30] In this method a velocity in the $x$-direction is added or subtracted to a particle velocity when it moves through the upper or lower boundary of the simulation box, which leads to a shear deformation. After a total deformation $\gamma = 1$ was reached (in $10^4$ steps using $\dot\gamma = 0.01$) the shear rate was put at zero, and the subsequent shear stress was followed in time. For $N_B/N_A = 1:8$ the same procedure was repeated with $\varepsilon_{AB} = 0$. This shows the stress decay when only entanglements are present. In all cases the shear stress $\sigma_{xy}$ is well fitted up to $t = 5000$ by

$$\sigma_{xy} = \sigma_0 \exp(-t/\tau) + G \qquad (17)$$

Here, the parameter $G$ is the residual stress at deformation $\gamma = 1$. If $G > 0$ the system has a finite plateau modulus that is roughly equal to $G$ (a non-linear stress-strain relation may cause a difference).

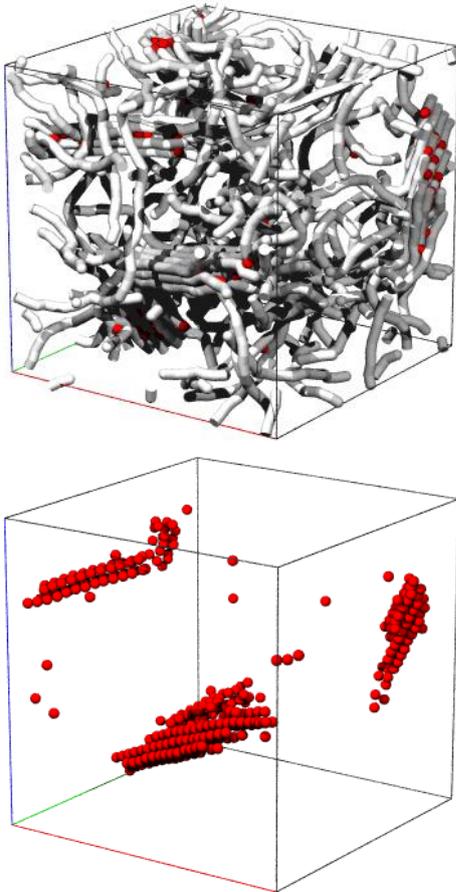

**Figure 17** Structure of mixed fiber/particle system for $N_B:N_A = 1:8$. Top picture shows the A fibers as wormlike chains, the bottom picture shows only the cross-linkers.

For the system without adhesive interaction ($\varepsilon_{AB} = 0$) we find $10^3 G = -0.05 \pm 0.05$, i.e. the residual stress vanishes. For the $N_B:N_A = 1:16$ and $N_B:N_A = 1:8$ systems we find respectively $10^3 G = 0.25 \pm 0.06$ and $10^3 G = 8.90 \pm 0.06$, i.e. finite values. These values are not very large, but generally a strong modulus is correlated to strong segregation. This example shows that it is possible to have a mixed gel with finite shear modulus, but without segregation.

Two pictures of the $N_B:N_A = 1:8$ system are shown in Figure 17. The top picture shows the full system, the bottom picture shows only the cross-linkers. It is observed that we have solid fiber bundle domains where several fibers are linked together, but these domains are connected together into a network by non-crystalline fibers. The right hand picture in Figure 17 shows that the cross-linkers migrate from one solid domain to another, similar to solid-vapour equilibrium. However, the loose cross-linkers are not in a true vapour phase but are associated to the fibers. Thus, cross-linkers migrate from one solid domain to another via the connecting fibers.

## VI. SUMMARY AND CONCLUSIONS

To structure liquids by associative semiflexible fibers we need to know how the network strength relates to fiber properties, and how to prevent syneresis at the same time. Fiber (bundle) networks are also abundant in many biological systems. Existing scaling relations for the rheology of semi-flexible polymers cannot be applied to fibers that form associative cross-links. Therefore insight needs to be gained by simulation. Networks of associative fibers and mixtures of fibers and colloidal particles are simulated.

First the phase diagram for adhesive colloidal particles is presented in terms of the particle elastic modulus and their surface energy. Next, the network mesh size and life time of the connections in associative fiber networks are studied, as these are key to network rheology. For non-associative fibers, the presence of explicit solvent is found to be important for the dynamics up to the collision time between fibers. After that, the mean square displacement with and without solvent superimpose after an appropriate time rescaling; the solvent only slows down diffusion. For associative fibers the same rate of slowing down is observed in the evolution when solvent is included. This suggests that we can safely leave out explicit solvent to gain simulation speed.

When fibers associate to each other, they form a network for sufficiently large fiber-solvent surface energy. If the surface energy is *above* the value where single particles crystallize the adhesion forces are large enough to alter the local direction of the fibers, and aligned dimers and trimers are formed early on. This process occurs in many places simultaneously, and one chain can be member of several fiber bundles, so that a network is formed immediately.

This network coarsens because non-associated chains join existing bundles, and because fiber bundles merge. This leads to syneresis and coarsening until the fiber bundles crystallize. Coarsening reduces chain curvature, and stops when the mean bending energy per bead becomes or order $kT$. Therefore, in this case the length of the network connections is roughly the persistence length, independent of surface energy.

If the surface energy is *below* the value where single particles crystallize, a network can still be formed but at a slower (reaction limited) rate. Loose (liquid-like) assemblies between chains are formed only when they happen to run more-or-less parallel. These assemblies grow by diffusion and aggregation and form a loose network, which sets in microphase separation. Only when the clusters crystallize, the coarsening process is stopped. In this case the length of the network connections is larger than the persistence length, and will depend on the value of surface energy.

When the particle elastic modulus and surface energy are such that associated chains form liquid fiber bundles, again a network is formed but this keeps on coarsening. Stress decays proportional to the logarithm of time, which suggests a stress induced coarsening process. Fibers are observed to reptate along the network connections where thick connections get fatter and thin connections break up. This is similar to the Plateau-Rayleigh instability of liquid tubes, but in this case the time scale of coarsening is most likely set by the diffusion of fibers, as long as the mesh size is shorter than the fiber length.

The solid networks obtained compare very well to experimental structures. For instance, the structures shown in Figure 8 are very similar to the artificial scaffolding structure of peptide-amphiphile fibers obtained by Hartgerink *et al*,[1] which resembles natural collagen networks and can act as a template for hydroxyapatite crystallisation. Also, the network structures found here are not too dissimilar from outer cell wall structures,[4] although cell wall fibers are ordered in one direction because of the proximity of the cell membrane.

Mixtures of fibers and colloid particles again form fiber bundle networks, but by choosing the colloid volume fraction below 1/7$^{th}$ of the fiber volume fraction (for fiber and colloid of the same diameter) swelling gels are obtained in simulation. This is in line with the theoretical predictions by Zilman and Safran.[17] The structure of the system is characterised by solid domains of colloid particles and fibers, in which several fibers are linked together. These fiber bundle domains are connected together into a network by non-crystalline fibers. The colloid particles migrate from one domain to another while remaining associated to non-crystalline fibers.

ACKNOWLEDGEMENTS
E.G. Pelan, S.D. Stoyanov, K.F. van Malssen and J. Nijsse are kindly acknowledged for stimulating discussions and advice.